# Fractal Geometry and Fractional Calculus for Integrative Morphological Mapping of Breast Cancer Complexity

Abhijeet Das[1,*], Ramray Bhat[1,2], Mohit Kumar Jolly[1]

[1]Department of Bioengineering, Indian Institute of Science, Bangalore – 560012, India

[2]Department of Developmental Biology and Genetics, Indian Institute of Science, Bangalore – 560012, India

[*]**Corresponding Author –** abhijeetdas@iisc.ac.in

**Abstract:** Breast cancer exhibits intricate morphological and dynamical heterogeneity across cellular, tissue, and tumor scales, posing challenges to conventional modeling approaches that fail to capture its nonlinear, self-similar, or self-affine, and memory-dependent behavior. Despite increasing applications of fractal geometry and fractional calculus in cancer modeling, their methodological integration and biological interpretation remain insufficiently consolidated. This review aims to synthesize these frameworks within an integrative morphological perspective to elucidate their collective potential for quantitative characterization of breast cancer complexity. Fractal geometry-based analyses quantify spatial and temporal irregularities along with spatiotemporal morphodynamics, while fractional calculus introduces non-local and memory-dependent formulations describing tumor growth. Together, these frameworks establish a mathematical link between fractal structure and fractional dynamics. Nevertheless, their application remains hindered by inconsistent methodologies and a lack of reproducible standards. This review consolidates existing evidence, delineates methodological interrelations between fractal geometry and fractional calculus, and outlines reproducibility requirements, including standardized preprocessing, parameter reporting, and benchmark datasets. Collectively, the findings emphasize that reproducible and biologically interpretable integration of these two approaches is fundamental to achieving clinically relevant modeling of breast cancer morphology and dynamics.

# 1. Introduction

## 1.1 Global Burden and Motivation

Breast cancer (BC) emerged as the leading global cancer in 2020 with approximately 2.3 million new cases, representing 11.7% of all cancer diagnoses and 7% of cancer-related deaths worldwide. It remains the most prevalent malignancy among women. [1] Early detection and prompt treatment constitute the most effective and economically viable control strategies. [2] Recognizing this imperative, the World Health Organization launched the Global Breast Cancer Initiative in March 2021, targeting a reduction in global BC mortality to 2.5% by 2040, with particular focus on supporting low- and middle-income countries through evidence-based technical assistance.

Despite advances in molecular diagnostics and gene therapy, these statistics highlight the limitations of existing BC research approaches, emphasizing the need for more effective diagnostic and predictive frameworks. This necessity motivates a shift in the interpretation of carcinogenesis, moving beyond strictly gene-centric, reductionist perspectives toward models that can capture morphological and dynamic complexity. This integrative view is schematically represented in Fig. 1, which illustrates the interconnection between biochemical cues, biomechanical constraints, and morphological alterations underlying cancer progression. [3]

Early detection of BC largely depends on recognizing structural and morphological alterations that precede plainly apparent clinical manifestations. Because biological systems exhibit highly correlated and organized architectures, their geometric irregularities carry information about underlying functional disruption. [3-6] Quantitative morphometric analysis based on measurable shape and structure parameters provides a means to detect these deviations objectively, complementing molecular and genetic diagnostics. Such structure-based indices can therefore improve early prognostic assessment by capturing the mesoscopic signatures of tissue disorganization described in the Tissue Organization Field Theory (TOFT) [6], bridging molecular data with morphological evidence of disease progression.

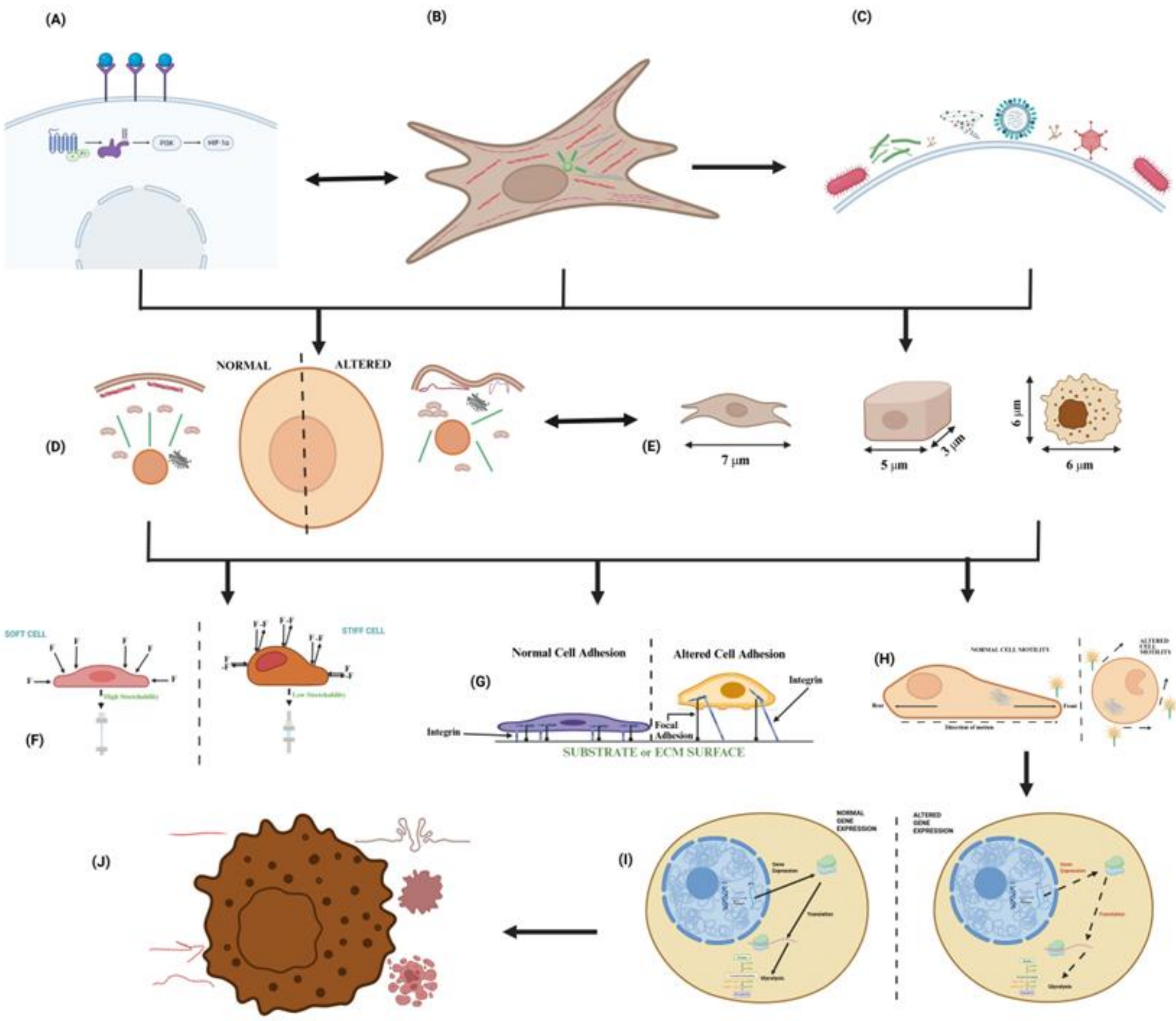

**Fig. 1** Schematic representation of the chemo-biomechanical pathways displaying the connection among shape features, cell biomechanics, gene-protein network, cytoskeleton, and disease state. Here, **(A)** Biochemical signals (intracellular or environmental), **(B)** Biophysical constraints/tensional homeostasis, **(C)** Foreign organisms and pathogens, **(D)** Structural changes induced cell membranes, cytoskeleton, and cytosol, **(E)** Variation in cell size and shape, **(F)** Variation in cell deformability, **(G)** Altered cytoadherence, **(H)** Altered cell locomotion and motility, **(I)** Altered cell function and gene expression, and **(J)** Cancerous state

## 1.2 The Need for Multiscale Quantification

In the early part of the 21st century, Hornberg *et al* [7] questioned whether the effects of genetic mutations could be meaningfully predicted if tumors and surrounding cells form a complex supracellular communication network. They argued that cancer research might progress more effectively when approached from a systems biology perspective rather than solely through molecular biology. This approach emphasizes a fundamental shift in philosophy beyond the reductionist, gene-centric framework by incorporating additional levels of organization such as cell and tissue morphology, microenvironmental dynamics, and spatial–temporal interactions. It accepts that tension between these two views remains a critical barrier to advancing future therapies. [8]

Modern biology has largely neglected organisms as integrated systems by failing to address problems across observational levels. The central dogma's reductionist focus on genes and molecular pathways has overshadowed higher-order organization, although biological systems exhibit correlated structures whose morphology cannot be ignored. [4] Thus, new frameworks are required to analyze the geometry and organization of living matter beyond molecular interactions.

At the mesoscopic scale, matter organization governs how forces, signals, and molecular gradients propagate through cells and tissues. [5] Physical form is not merely a genomic output but a product of internal and external constraints [9], and in this context, while the genome represents a digital core of information [10], it does not fully account for the emergent complexity or tissue-level behavior of biological systems. [11] Notably, empirical observations show little direct correlation between genomic and morphological complexity [12], indicating that biological complexity cannot be localized solely at the genomic level. Systems biology, therefore, advocates quantifying and interpreting multi-scale patterns of morphological organization and their dynamic regulation. [8] Morphological analysis thus becomes a powerful means to probe system-level states and emergent behaviors.

Accordingly, understanding cancer requires mathematical and computational tools capable of capturing multi-scale geometry and non-linear dynamics that traditional approaches cannot represent adequately.

## 1.3 Theories and Mathematical Frameworks for Cancer Complexity

Two major paradigms frame carcinogenesis. The Somatic Mutation Theory (SMT) posits that cancer arises from deterministic mutations leading to progressive genetic alterations. [13] This Darwinian micro-evolutionary model [14] assumes that genes govern cell functions linearly, an assumption now recognized as insufficient. [5] In contrast, the TOFT proposes that cancer emerges from the disruption of cell-to-cell junctions, morphostatic gradients, and tensional homeostasis within tissues, representing a breakdown of architecture. [6] This view is supported by the emergent theory of carcinogenesis [15], which highlights the importance of mesoscopic observations and quantitative morphological measures.

Traditional models based on linear or non-linear differential equations (LDEs and NLDEs) have successfully described phenomena such as action potentials and feedback control in physiology [16], yet they struggle to capture the multi-scale and memory-dependent dynamics of living systems. According to us, a possible way to bridge molecular and organ-level behaviors is by developing multi-scale strategies, including probabilistic, fractal, and chaotic approaches [17], along with integrating non-integer order derivatives. Specifically, Fractal Geometry (FG) for quantifying morphological complexity, whereas Fractional Calculus (FC) for modeling non-local temporal and spatial dependencies through fractional derivatives (e.g., Caputo, Riemann–Liouville). [18-20] Together, FG and FC offer complementary frameworks for describing the geometry and dynamics of cancer as a complex biophysical system. This conceptual transition from linear deterministic systems to complex, scale-invariant, and fractal frameworks is illustrated in Fig. 2, highlighting why modeling cancer dynamics demands approaches beyond traditional Euclidean or linear paradigms. [21]

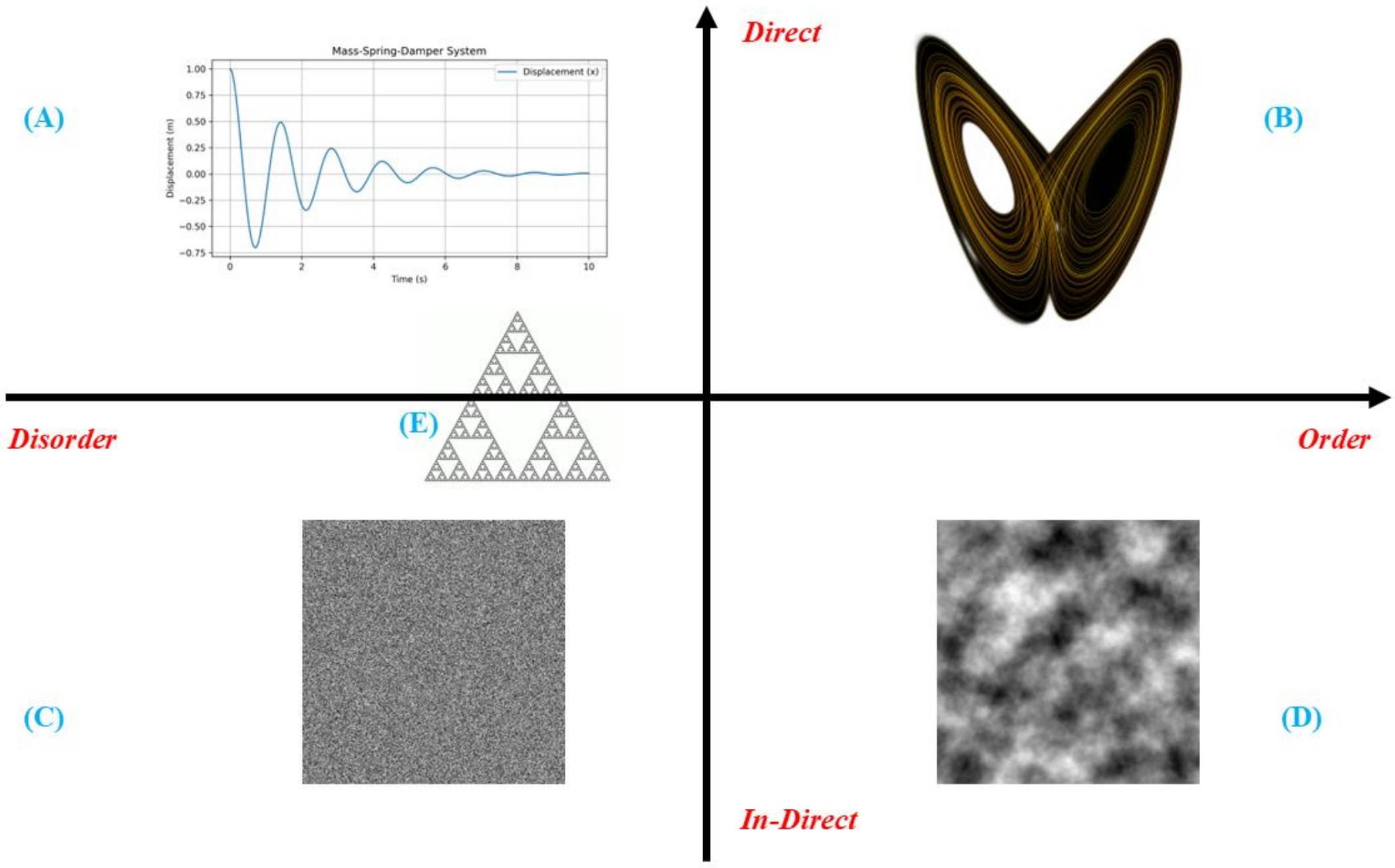

**Fig. 2** Relationship between models/frameworks used for investigating complex dynamical systems. Here, **(A)** Linear time-invariant causal system is represented by the mass-spring-damper system, **(B)** Chaotic system, visually highlighted by the Butterfly effect, **(C)** White noise, **(D)** Turbulence, visually represented by the Perlin noise, and **(E)** Fractals, graphically represented by the Sierpienski triangle

FG and FC are inherently connected through their shared capacity to describe systems exhibiting scale invariance, non-locality, and memory. In biological contexts, the fractal organization of tissues reflects structural irregularities and long-range spatial correlations, while fractional derivatives mathematically can capture the corresponding temporal and spatial memory effects within dynamic processes. Within the framework adopted here, these complementary approaches together enable quantitative characterization of both morphological complexity and the underlying dynamical behavior of BC systems. Accordingly, this review focuses on integrating applications of fractal geometry and fractional calculus in breast cancer research to advance multiscale understanding and predictive modeling.

### 1.4 Research Hypothesis

Studying BC as a multiscale, emergent property of cellular/tissue systems within an integrative morphological mapping, quantifying morphological complexity through FG, and modeling biological dynamics using FC may meaningfully enhance current BC research approaches. While LDEs and NLDEs have been valuable for modeling tumor dynamics and pharmacokinetics [16], they remain limited in representing long-term memory and non-local effects (hereditary effects, anomalous transport). In addition, micro-simulation models, e.g., agent-based and cellular automata models, offer detailed cell-level resolution but often suffer from computational intractability when scaled to simulate tissue-level organization and dynamics. [22] We argue that FC-based models uniquely incorporate these properties, allowing the system's present state to depend on its entire history.

This remaining part of the review is organized as follows: Section 2 introduces the fundamentals and key algorithms of FG for morphometric quantification; Section 3 reviews applications of FG in BC research across different modalities; Section 4 presents mathematical basics of FC and its derived modeling approaches, along with conceptual linkage between geometry and dynamics; Section 5 discusses prospects; and Section 6 summarizes key conclusions.

## 2. Fractal Geometry for Morphometric Analysis

### 2.1 Conceptual Overview: Why Fractals for Cancer Morphology

Classical Euclidean geometry effectively represents only regular objects with integer dimensions (1, 2, or 3) and fails to characterize the irregular, hierarchically organized forms typical of natural and biological systems, including malignant tumors. The irregularity and heterogeneity of tumor cells and tissue architecture highlight the limitations of Euclidean measures.

Benoît Mandelbrot introduced FG as a mathematical framework capable of quantifying complex morphologies that exhibit self-similarity and scale invariance. The term fractal (from Latin frāctus, "broken" or "fragmented") captures this essential property. Fractals possess a Hausdorff–Besicovitch dimension greater than their topological dimension (TD), although space-filling curves such as those of Hilbert, Peano, or Koch demonstrate equality between the two dimensions. A mathematically comprehensive discussion of such curves can be found in Sagan's work. [23]

An object or a pattern is identified as fractal when it exhibits one of the following forms of self-similarity (Fig. 3):

1) Exact self-similarity – The object maintains an identical structure or pattern across all scales (Fig. 3(a)).

2) Quasi self-similarity – Approximate repetitions of the overall pattern are observed at various scales, albeit with distortions (Fig. 3(b)).

3) Statistical self-similarity – Patterns repeat across scales, but only statistically, maintaining scale-invariant statistical properties (Fig. 3(c)).

4) Multi-fractals – More than one scaling rule or $FD$ is present across different regions (Fig. 3(d)).

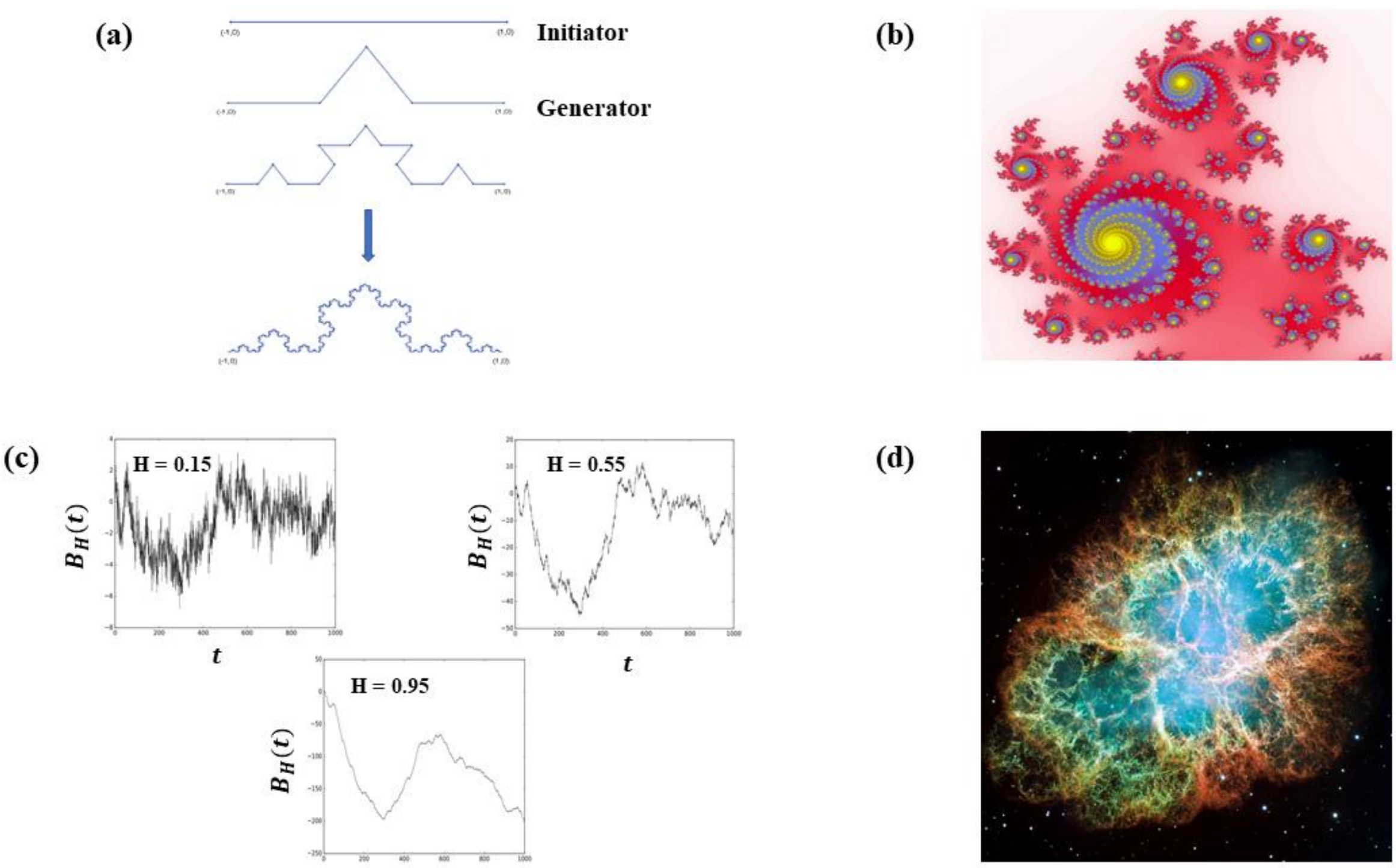

**Fig. 3 (a)** Koch curve - Initiator step consists of having a straight line with unit length. In the generator step, the line is divided into three fragments, with the middle third as the base, constructing an equilateral triangle with the removal of the base. The resulting figure consists of $4^1$ line segments, each with $\frac{1}{3^1}$ length, while the total length is $\left(\frac{4}{3}\right)^1$. In the next step, the middle third of each line segment is the base equilateral triangles are constructed with the removal of bases. The resulting figure will have $4^2$ line segments, with the length of each being $\frac{1}{3^2}$, such that the total length becomes $\left(\frac{4}{3}\right)^2$. Successive iterations result in the curve progressively winding with approaching the limiting curve, i.e., the Koch curve, **(b)** Julia set corresponding to a complex equation $0.355 + 0.355\text{i}$, **(c)** Time trace of particle's position executing fractional Brownian motion with varying Hurst exponent, consequently, displaying persistent, anti-persistent, and no memory-effect, and **(d)** An astrophysical multi-fractal structure-Crab nebula, reported to display multi-scaling characteristics, resulting in its structural complexity

A distinctive consequence of fractality is that measured quantities such as length, area, or volume depend on the scale of measurement. For example, the British coastline example illustrates that measured length increases as the ruler or scale of resolution becomes finer. Thus, fractal structures lack a fixed metric; their geometrical measures are inherently scale-dependent.

The Fractal Dimension (FD) quantifies this scaling, linking the measured property to the observational scale. In spatial data, FD expresses morphological complexity or surface roughness; in time series, it represents the degree to which fluctuations increase the path's complexity. Higher FD values denote greater irregularity and long-range correlations, while lower FD values correspond to smoother configurations. This relationship connects to the Hurst exponent ($\alpha$) from the equation $FD = TD + 1 - \alpha$ where, $\alpha < 0.5$ implies anti-persistence (or, negative correlation), $\alpha > 0.5$ persistence (or, positive correlation), and $\alpha = 0.5$ randomness or Brownian motion-like behavior, as shown in Fig. 3c. [24-26]

Various algorithms have been developed to compute FD, including the Box-Counting Algorithm, Higuchi's Algorithm, and Power Spectrum Density (PSD) analysis. Yet FD alone cannot capture all aspects of structural complexity. [27, 28] Consequently, it is frequently complemented by fractal's Lacunarity (LC), a geometric descriptor of texture and spatial heterogeneity that assesses void or gap distribution within the structure along with rotational or translational invariance. [29-31] High LC values signify greater heterogeneity, whereas low LC values indicate spatial uniformity.

The Box-Counting method can be adapted to compute LC by evaluating the variance of pixel occupancy across scales, rather than counting occupied boxes. Together, FD and LC describe mono-fractal properties, assuming a single scaling law or behavior. However, biological structures may exhibit multi-fractal behavior with multiple scaling parameters at different scales; therefore, multi-fractal analysis should be accompanied by mono-fractal assessments to avoid ambiguity. To this end, Multifractal Detrended Fluctuation Analysis (MFDFA) has been widely applied to detect multi-fractal signatures of morphological complexity in diverse systems, including cancer. [32-34] Motivatingly, we have discussed the Two-Dimensional Multifractal Detrended Fluctuation Analysis (2D-MFDFA) [26, 32], for extracting multi-fractal parameters and spectrum, respectively, from time series data.

## 2.2 Core Methods and Algorithms

### 2.2.1 Fractal Dimension

#### 2.2.1.1 Box-Counting Algorithm

In general, this method is equivalent to partitioning space into $n$-dimensional boxes of defined side length. In other words, the approach utilized measures the characteristics/features of objects/systems at different scales, plots a graph between feature versus scale, and fits a least-squares regression line where the slope gives the FD of the object or system. The Box-Counting algorithm, in particular, utilizes a set of boxes with a defined side length $(\varepsilon)$, which are then used to make a grid and placed over the object of interest. Subsequently, the number of boxes required to completely cover the object $N(\varepsilon)$ is counted. The process is repeated with boxes of different sizes, and a graph is plotted in log-log scale between $N(\varepsilon)$ and $1/\varepsilon$ over the linear scaling range to compute the FD (Fig. 4(a) and 4(b)). The algorithm thereby measures how morphological features occupy space at progressively finer scales. Mathematically, it is represented by Eq. 1:

$$FD = \lim_{\varepsilon \to 0} \frac{\log N(\varepsilon)}{\log\ (1/\varepsilon)} \tag{1}$$

However, the direct approach finds limitations in the utilization for grayscale images (e.g., scanning electron microscope images or mammograms). In this regard, the Differential Box-Counting (DBC) algorithm was proposed [35] for the FD of textured images. The approach is based on the 3D representation of a 2D grayscale image, where the third axis or coordinate represents the gray-level intensity of each pixel. Nonetheless, the algorithm does not take into consideration the relative changes in gray-level, contributing to the textural complexity. Consequently, the Relative DBC (RDBC) method was developed by Jin *et al* [36]; however, instead of utilizing the absolute box $(i, j)$ height difference, it computes the relative difference for each grid from Eq. 2:

$$r_g(i, j) = M_g(i, j) - m_g(i, j) \tag{2}$$

Here, $M_g(i,j)$ and $m_g(i,j)$ represents the maximum and minimum gray-level values in the grid, respectively. Subsequently, the number of occupied boxes in each grid is computed using Eq. 3:

$$N(\varepsilon) = \sum_{i,j} \left\lceil \frac{k \times r_g(i,j)}{\varepsilon} \right\rceil \tag{3}$$

Here, $k = \frac{M}{G}$ is the scaling factor in the gray-level or third axis, where, $M$ is the image dimension and $G$ is the maximum gray-level, and $\lceil \ \ \rceil$ is the ceiling function. The FD is then computed from the above-mentioned linear-regression approach. Nevertheless, here the FD is concerning a specific direction; thus, the investigated patterns are not self-similar but self-affine in nature.

### 2.2.1.2 Higuchi's Algorithm

The algorithm was first utilized to compute the FD of an irregular time series in the time domain itself. [37] Here, a discrete series of data points is constructed with $N$ total data points consisting of values at regular intervals. Then, from the single series of data points, new $k$ sub-sequences $\mathrm{S_m}(k)$ is constructed (Eq. 4) where $\mathrm{m} = 1, 2, \dots \dots, \mathrm{d}$ represents the initial time, and $k$ is the time interval with the property $1 \le \mathrm{k_{max}} \le \left\lceil \frac{\mathrm{N}}{2} \right\rceil$. [26]

$$S_m(k) : x(m), x(m+d), x(m+2d), \dots \dots \dots x\left(m + \left\lfloor \frac{N-m}{k} \right\rfloor k\right) \tag{4}$$

Subsequently, the length $\mathrm{L}_m(k)$ is computed using Eq. (5):

$$L_m(k) = \frac{1}{k}\left\{\left(\sum_{i=1}^{\left\lfloor \frac{N-m}{k} \right\rfloor} |x(m+\mathrm{i}k) - x(m+(\mathrm{i}-1)k|\right)\frac{N-1}{\left\lfloor \frac{N-m}{k} \right\rfloor k}\right\} \tag{5}$$

Here, $m$ and $k$ are integers, $\lceil \ \ \rceil$ and $\lfloor \ \ \rfloor$ are ceiling and floor functions, respectively. The length $L_m(k)$ represents the normalized sums of the absolute value of the difference in pairs of data points situated at $k$ distances from the initial point/time $m$.

For each sub-sequence $k$, the mean length is calculated using Eq. 6:

$$L(k) = \frac{1}{k}\sum_{\mathrm{m}=1}^{k} L_m(k) \tag{6}$$

Finally, the FD is computed from the least-squares fit of the plot between $L(k)$ and $k$ on a double logarithmic scale. It is to be noted that the FD computed from Higuchi's algorithm always lies in the closed interval [1, 2], where smooth curves like sine and cosine display $FD = 1$, while randomly distributed or stochastic curves will show $FD = 2$. Nonetheless, there exists an exception in the case where all the data points possess equal value; subsequently, $L_m(k)$ becomes zero, resulting in $FD = 0$. [38]

### 2.2.1.3 Power Spectrum Density

It is based on converting the investigated image to the frequency domain, where the object's/system's features are described by wave numbers using the Fourier transform. In this regard, the PSD method brings to light the wavelengths contributing to the investigated feature. Mathematically, this method is the Fourier transform of the autocorrelation function of signals composing the object and identifies the present spatial frequencies within a range of wave-vectors. [39] In image analysis, the power spectrum is given by Eq. 7:

$$P(k_x, k_y) = c|\vec{k}|^{-\beta} \tag{7}$$

The least square approximation (Eqn. 8) gives the scaling exponent by,

$$\beta = \frac{N \sum_{ij} \log_e|k_{ij}| \log_e P_{ij} - \sum_{ij} \log_e|k_{ij}| \sum_{ij} \log_e P_{ij}}{N \sum_{ij} (\log_e|k_{ij}|)^2 - (\sum_{ij} \log_e|k_{ij}|)^2} \tag{8}$$

Here, k, N, i, and j represents the wave-vectors, number of data points, indices in the horizontal and vertical directions, respectively. Notably, the FD for self-affine objects is computed using Eq. (9): [38]

$$FD = \frac{8-\beta}{2} \tag{9}$$

## 2.2.2 Lacunarity

### 2.2.2.1 Gliding Box-Counting Algorithm

The pictorial representation of the difference between box-counting and gliding box-counting methods is shown in Fig. 4(c). In this approach, the object's image(s) are converted into binary format following the condition in Eq. 10:

$$g(x, y) = \begin{cases} 1 \text{ } if \text{ } k(x, y) \geq h^* \\ 0 \text{ } otherwise \end{cases} \tag{10}$$

Here, $k(x, y)$ signifies an individual object's feature with $x$ and $y$ pixel's coordinates and $h^*$ is the threshold value. The distribution of lacunar pixels in the object image is evaluated using the gliding-box algorithm. In this method, the number of boxes with length $l$ and $p$ lacunar pixels is represented by the frequency distribution $n(p, r)$. The probability distribution is computed from Eq. (11) as,

$$P(p, r) = \frac{n(p,r)}{(\Delta_a - r + 1) \cdot (\Delta_b - r + 1)} \tag{11}$$

Here, the quantity $(\Delta_a - r + 1) \cdot (\Delta_b - r + 1)$ is the total number of boxes corresponding to the image's height and base, respectively. Subsequently, the LC is computed using Eq. 12:

$$LC(p, r) = \frac{\sum p^2 \cdot P(p,r)}{[\sum p \cdot P(p,r)]^2} \tag{12}$$

Finally, the coefficient at each scale is computed from the curve fitting of the decrement in LC with the increment in separation between the pixels $(r)$ in the log-log scale. [40]

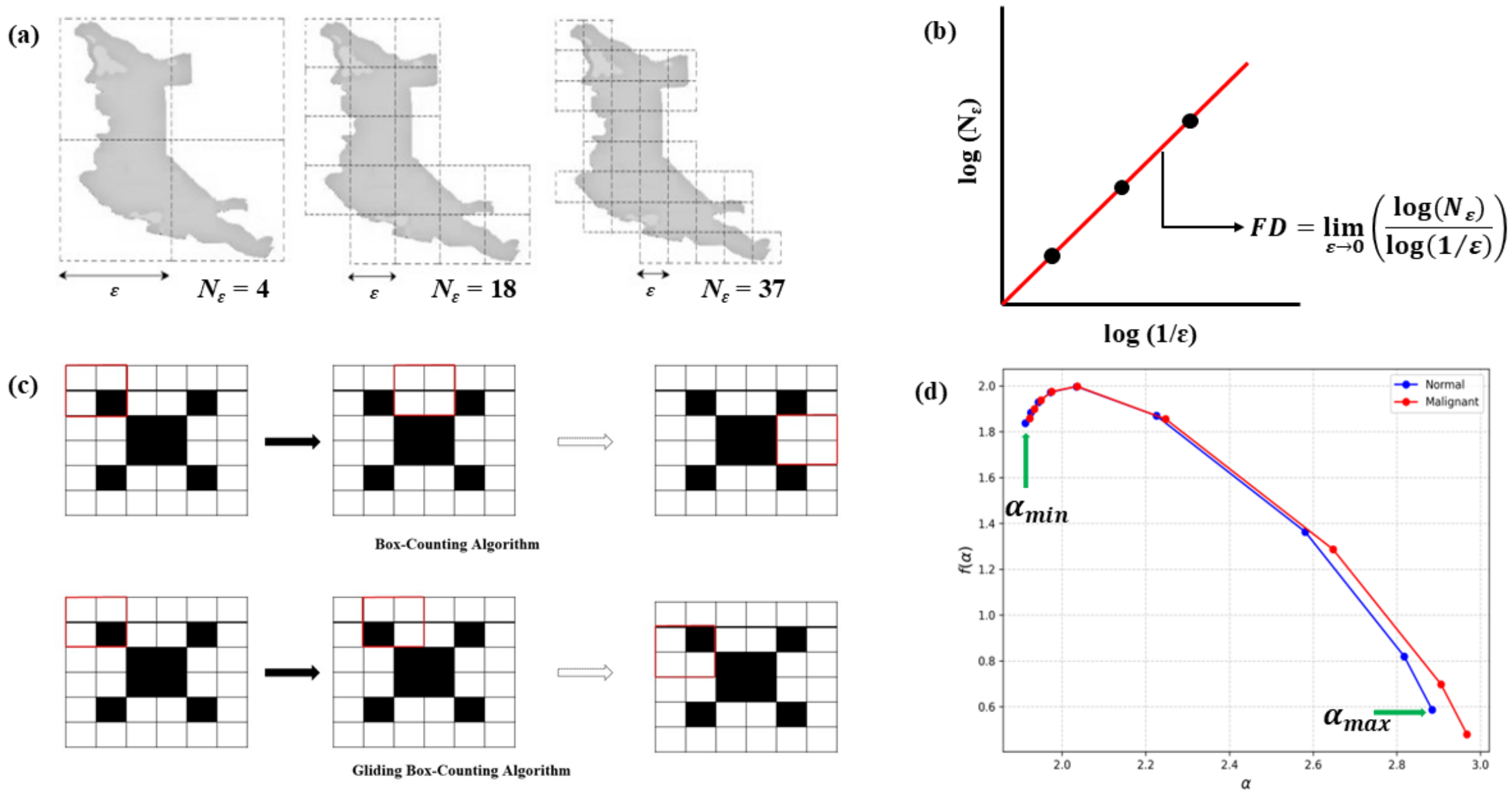

**Fig. 4** Pictorial representation of **(a-b)** box-counting algorithm for measuring the fractal dimension of an irregular/complex structure and **(c)** difference between box-counting and gliding box-counting algorithm for computation of fractal dimension and lacunarity, respectively, on a 6 × 6 binary image and **(d)** Averaged multi-fractal spectrums exhibiting the difference in multifractality strength between normal and malignant mammograms from the Chinese Mammography Dataset. The detailed study, including analysis of individual mammogram, is reported in ref. [41].

### 2.2.3 Succolarity

#### 2.2.3.1 Differential Box Counting

Succolarity (SC) is a fractal-based texture descriptor that quantifies the degree of percolation within an image structure, reflecting how easily a virtual fluid could pass through the spatial arrangement of pixels. It complements FD and LC by incorporating directional and connectivity-based information related to image texture and flow pathways, rather than self-similarity or gap distribution. While FD and LC capture morphological complexity and heterogeneity, respectively, SC evaluates the permeability and flow capacity of the texture pattern.

Notably, SC is generally utilized for binary images where all black pixels are considered as empty spaces, and white pixels as obstacles. The image is divided into equal-sized boxes $BS(k)$, where $k$ is a divisor of the image dimension, similar to the box-counting method. For each box, the occupation percentage $OP(BS(k))$ is calculated, and the pressure above the centroid of the box is used to evaluate the flow contribution according to Eq. 13:

$$\sum_{k=1}^{n} OP\big(BS(k)\big) \times PR\big(BS(k)\big), pc \tag{13}$$

Considering the dimensionless feature of FD and LC, respectively, the dimensionless succolarity is obtained using Eq. 14, considering that the image was totally flooded by the fluid or the image is totally filled with black pixels. [42, 43]

$$SC(BS(k), direction) = \frac{\sum_{k=1}^{n} OP\big(BS(k)\big) \times PR\big(BS(k)\big), pc}{\sum_{k=1}^{n} PR\big(BS(k)\big), pc} \tag{14}$$

Here, $PR\big(BS(k)\big), pc$ represents the pressure, where $pc$ is the position on $x$ or $y$ of the centroid of the box on the scale of pressure applied to the box, $direction$ represents the inlet direction (top-to-bottom, bottom-to-top, left-to-right, and right-to-left) of the virtual fluid. Nevertheless, here pressure represents a notional driving potential applied across the binary image to simulate directional percolation, while direction defines the axis along which this flow is evaluated. It serves as a mathematical weight quantifying how easily a virtual fluid could traverse the open pathways in an image. Thus, evaluating SC across four orthogonal directions can capture structural anisotropy and connectivity, offering a fractal analogue of permeability in biological textures.

Beyond measures of morphological complexity and connectivity, the inherent heterogeneity of biological image features, such as textures, can be quantified through multifractal analysis, which captures variations in scaling behavior across different spatial resolutions.

### 2.2.4 Multi-fractal and Multi-fractal Detrended Fluctuation Analysis

Multi-fractal Analysis (MFA) extends classical fractal geometry or mono-fractal geometry by characterizing the presence of multiple scaling exponents within a structure, thereby quantifying spatial heterogeneity at different scales. Unlike the mono-fractal description provided by a single FD, MFA identifies how the distribution of local singularities evolves with scale. According to Das *et al* [41], multifractal measures provide a more comprehensive quantification of textural complexity in gray-scale images and the coexistence of diverse growth patterns and organizational scales, subsequently, capturing variations that cannot be resolved by global fractal parameters alone.

Mathematically, MFA relies on the partition function formalism, which quantifies the probability distribution of masses (e.g., pixel intensities) across scales. For an image divided into boxes of size $\varepsilon$, the partition function is expressed from Eq. 15:

$$Z(q, \varepsilon) = \sum_{i=1}^{N(\varepsilon)} [p_i(\varepsilon)]^q \tag{15}$$

where $p_i(\varepsilon)$ represents the normalized measure (mass or gray-level intensity) within the $i^{th}$ box, and $q$ is the moment order. For a fractal object or for a scale-limited fractal (e.g., natural or biological object), the partition function follows a scaling or power-law behavior with box size that follows Eq. 16:

$$Z(q,\varepsilon) \sim \varepsilon^{\tau(q)} \quad (16)$$

where $\tau(q)$ denotes the mass exponent function. The local singularity strength $\alpha = \frac{d}{dq}\tau(q)$ and the multifractal spectrum $f(\alpha)$ are related by the Legendre transform, given by Eq. (17):

$$f(\alpha) = q\alpha - \tau(q) \quad (17)$$

and quantifies the distribution of local scaling exponents (Hölder exponents). A wider multifractal spectrum, computed as the difference between the maximum ($\alpha_{max}$) and minimum ($\alpha_{min}$) values of the local singularity strength or global singularity (Hausdorff fractal dimension) determine the strength of multifractality, and subsequently, indicate a broader range of scaling behaviors, hence greater heterogeneity within the structure. [26, 44]

While the MFA formalism defines the theoretical framework, its application to biomedical images can introduce analysis bias by background gradients or non-stationary intensity fields. To address this, the Multi-fractal Detrended Fluctuation Analysis (MFDFA) algorithm implements the MFA concept computationally, providing robust estimation of multifractal parameters in 2D non-stationary data. In this method, the overall fluctuations in a system for the $q^{th}$ order moment are given using Eq. 18:

$$F_q(n) = \left(\frac{1}{M_n N_n}\sum_{k_1=1}^{M_n}\sum_{k_2=1}^{N_n}\left(F(k_1,k_2,n)\right)^q\right)^{1/q} \quad (18)$$

Here, $q$ can take any integer value, $M_n \times N_n$ represents the disjoint segments of the feature with equal sizes $n \times n$. The least squares fit of the graph between $F_q(n)$ and $q$ in a log-log scale gives the scaling of the fluctuation function, known as the Generalized Hurst exponent $h(q)$. The mass exponent is computed from $h(q)$ using Eq. 19:

$$\tau(q) = qh(q) - FD \quad (19)$$

In Eq. (19), FD represents the fractal dimension of the geometric support of the multi-fractal measure. In addition, it is noteworthy that the parameters $h(q)$ and $\tau(q)$ displays a non-linear behaviour in the case of the presence of multi-fractal characteristics in a system.

For 2D or quasi-3D systems, in the 2D-MFDFA, the embedding space is partitioned into $N(\varepsilon)$ boxes of a size $\varepsilon$. Following Eq. (15) and assuming $p_i \sim \varepsilon^{\alpha}$ in the limit $\varepsilon \to 0$, the number of boxes with a scaling exponent between $\alpha$ and $\alpha + d\alpha$ is nearly equivalent to $\varepsilon^{-f(\alpha)d\alpha}$. Thus, Eq. (15) modifies into Eq. (20)

$$Z(q,\varepsilon) = \int \varepsilon^{-f(\alpha)+\alpha q} d\alpha \tag{20}$$

The smallest value of $\varepsilon^{-f(\alpha)+\alpha q}$ satisfies Eq. 21 and is given as,

$$\frac{d}{d\alpha}[\alpha q - f(\alpha)]_{\alpha_m} = 0 \tag{21}$$

$$\Rightarrow f'(\alpha_m) = q \text{ and } Z(q,\varepsilon) = \varepsilon^{\tau(q)}$$

Thus,

$$\frac{d\tau}{dq} = \alpha_m + \frac{d\alpha_m}{dq} q - f'(\alpha_m)\frac{d\alpha_m}{dq} = \alpha_m$$

$$\Rightarrow f(\alpha_m) = q\frac{d\tau}{dq} - \tau \tag{22}$$

The continuous curve traced by $f(\alpha_m)$ with $\alpha_m$ with a variation of $q$ in the interval $[-\infty, \infty]$ is known as the multifractal spectrum (Fig. 4(d)).

## 2.3 Strengths, Limitations, and Best Practices

Box-counting and Gliding Box methods provide direct spatial quantification and are computationally simple but may be sensitive to thresholding and noise. Higuchi's Algorithm is appropriate for analyzing one-dimensional profiles or contour-derived signals, avoiding image segmentation requirements. PSD analysis offers efficient frequency-domain characterization but presumes signal stationarity. MFDFA addresses non-stationarity and captures heterogeneity by analyzing a continuum of scaling exponents.

When combined, these techniques yield complementary insights, i.e., FD captures global irregularity, LC describes spatial heterogeneity, succolarity quantifies connectivity or percolation-based trends, and MFA/MFDFA reveals the distribution of local singularities. Nonetheless, consistent preprocessing (e.g., binary (thresholding, segmentation) or grayscale conversion (histogram normalization), preservation of aspect ratio, padding to maintain image dimensional consistency for ROIs of different dimensions, morphological operations), appropriate scaling-range selection, and validation across independent samples and different imaging modalities, along with explicit description, are crucial for reproducibility and biological interpretation. Also, acquisition parameters, e.g., device/model, imaging protocol, resolution, voxel/pixel size, and patient or sample metadata, e.g., number of cases, inclusion/exclusion criteria, clinical labels, follow-up, should be reported. Applying both mono- and multi-fractal analyses ensures robust quantification of tumor morphological complexity.

## 3. Applications of Fractal Geometry in Breast Cancer Research

The application of FG to BC has emerged as a robust approach for quantifying the morphological complexity associated with tumor growth and progression. By providing numerical descriptors, e.g., FD and LC, fractal analysis allows objective differentiation of morphological patterns across cellular, tissue, and imaging scales. This section presents the major domains of FG application in BC research, from cytological and histological image studies to radiological modalities. Additionally, we discussed a stochastic-based approach, i.e., dynamic scaling theory, combining evolution in tumor morphology with the spatiotemporal dynamics.

### 3.1 Cytological and Histological Image Studies

Early investigations in cytological and histopathological domains established the foundational role of fractal analysis in quantifying morphological disorder in breast tissues. These studies focused on cellular and nuclear architectures visualized in Hematoxylin and Eosin (H&E)-stained slides or cytological smears, demonstrating that malignant transformation is accompanied by increased geometric irregularity and self-similarity.

Rizki and Bissell [45] observed that malignant neoplasms, such as invasive BC, often lack structural organization and functional coordination with surrounding normal tissues. They argued that this irregularity leads to an increase in morphological complexity at the subcellular, cellular, and multicellular levels, and quantifying this complexity could correlate with patient outcomes. In this context, Tambasco *et al* [46] computed the FD of segmented histological structures from pan-cytokeratin-stained breast tissue microarrays using the box-counting algorithm. Analyzing samples from 379 patients, they demonstrated that increased epithelial morphological complexity or scale-invariant irregularity of epithelial architecture significantly correlated ($p < 0.001$) with disease-specific and overall survival. Specifically, the morphological complexity reflected the disorganization at the sub-cellular (keratin distribution and nuclear shape), cellular (cell contours), and multicellular (glandular formation) levels. The prognostic value of BC tumours was assessed using FD and LC parameters on a group of 40 low-risk patients who did not undergo any systematic treatment. [47] The samples were collected from surgically removed tissue sections and were subsequently stained with H&E dyes. The FD and LC were computed from the digitally photographed tissue sections (x400 magnification) using the box-counting algorithm. They observed, via comprehensive statistical evaluation, a correlation between FD and LC with tumours' biological properties, thus offering a promising and economical strategy for assessing the risk of distant metastasis independently of molecular biomarkers. Additionally, the fractal characteristics of native tumour histology were reported as effective prognostic markers with a long median follow-up period of 5 months. Yokoyama *et al* [48] studied the implication of irregularity in cell cluster edge-shape in cytological diagnosis of BC using image analysis. They investigated the edge-shape irregularity in cell clusters as a diagnostic criterion for differentiating between benign and malignant tumours via comparison of breast tumours demonstrating weak cellular atypia in low-grade invasive ductal carcinoma (IDC) using box-counting computed FD, along with 8 other parameters. Fine needle aspiration specimens of tumours were collected from 37 patients (16 low-grade IDC and 21 benign fibroadenoma (FA)), and 740 cell clusters were examined based on grouping into three types, viz., IDC clusters, FA with irregular clusters, and FA with regular clusters. Interestingly, they reported the average cluster size area in FA with irregular clusters to be approximately 3 times larger than that of IDC clusters. Consequently, they emphasized the focus on irregularities of cluster edge-shape in the case of differentiating between IDC and FA for accurate diagnosis. In a similar consideration, the effectiveness of FD for differentiating between normal and cancerous breast and 3 other cells' nuclear texture, was studied from the image intensity using the fractional Brownian function, from the assumption that the cell nucleus displays a fractal property. [49] The results suggested the significance of the scaling range in fractal investigation since the range over which cancerous cells displayed fractal properties was considerably larger than normal cells. In addition, they argued that the Hurst exponent can be used to classify the cells because of the way it is computed, i.e., for cancerous cells, it was computed over the nucleus since the nucleus almost entirely covers the cell while for normal cells, it is computed over the nucleus and cytoplasm.

## 3.2 Radiological Image Studies

The principles of fractal geometry have also been extended to radiological imaging modalities, enabling the quantification of lesion boundary roughness and internal texture at larger anatomical scales. In mammography, ultrasound, and magnetic resonance imaging (MRI), FG-based analysis has been applied to evaluate whether fractal metrics can enhance lesion characterization and diagnostic accuracy.

In radiological imaging modalities like mammograms, microcalcifications are small specks of calcium deposits and are often a finding in early stages of BC. In addition, they are radio-opaque and thus appear white in mammograms. In this context, a hierarchical interaction between morphological descriptors and parenchyma FD (computed using the box-counting algorithm) was studied for discriminating between benign and malignant categories using digital mammography. [50] The study included 31 patients with microcalcification, and confirmed from stereotactic biopsies, and were classified according to the Breast Imaging-Reporting and Data System (BI-RADS), along with parenchyma FD and biopsy size. The results implied the possible usage of quantitative shape evaluation and parenchyma FD for promising prediction of BI-RADS score. In addition, the lesions' area and parenchyma FD exhibited a complex distribution for malignant breast microcalcifications, which was in agreement with the observed qualitative morphological patterns. Additionally, a hybrid feature extraction method from mammograms to detect and classify microcalcification, architectural distortion, breast masses (or space-occupying lesions), and bilateral symmetry (or asymmetry of breast parenchyma between two sides) was proposed based on multifractal analysis (Renyi FD spectra), directional and morphological analysis, and Gabor filters. Here, the regions of interest (ROI) were identified using intuitionistic fuzzy clustering, and feature classification was done using a self-adaptive resource allocation network. The proposed method was implemented on images taken from open-access databases- Digital Database for Screening Mammography (DDSM) and Mammographic Image Analysis Society (MIAS), and subsequently, exhibited accuracy (sensitivity) of 93.75% (0.93) and 94.72% (0.92) for DDSM and MIAS, respectively. [51]

It is also noteworthy that distortion of architecture in breast parenchyma, including radiation of spiculation from a point and focal distortion at the parenchyma edge without an increase in breast mass density, is the third most common indication observed in the mammographic signature of nonpalpable BC. [52] This architectural distortion can appear in the initial stages of BC; however, owing to its ability to mimic normal breast tissues, its presence is often missed during screening and is reported to be one of the most common factors in false-negative cases. [53] Regarding the aforementioned, Banik *et al* [54] aimed to develop a computer-aided diagnostic technique for the detection of architectural distortion in prior mammograms of interval-cancer cases utilizing Gabor filters, linear phase portrait analysis, $FD$ from PSD, and the angular spread of power in the frequency domain. They used 1745 digitized mammograms of 170 patients obtained from the Alberta Program for the Early Detection of BC. Prior mammograms of interval-cancer cases (106 nos.) were identified by a radiologist, and subsequently, two categories, viz., visible architectural distortion (38 nos.) and questionable/invisible architectural distortion (38 nos.) were made for the study. The outcome suggested that a combination of FD and angular spread of power can be used to detect the subtle signatures of architectural distortion in mammograms. A different study, although carried out on a small set of 19 mammograms, reported the combination of $FD$ and $LC$, giving a prediction accuracy of 90% in the detection of architectural distortion. [55]

The fractal $SC$ parameter has recently been prominently utilized in the analysis of textures in different systems [56, 57], although it has found limited usage in biology as of now. Nonetheless, $SC$ has been utilized in the classification of BC masses collected from the MIAS and INbreast datasets. According to the authors, this parameter measures the roughness of the contours and, in combination with FD and LC, can effectively differentiate between normal, benign, and malignant tumors. However, no clear correlation between FD and SC was observed in this study. [42] Nonetheless, we introduced a modified SC-based but direction-independent measure, namely the succolarity reservoir, to account for latent-connectivity in tissue architecture from mammograms. [41] The measure was observed to hold statistical significance, in addition to $FD$ and multifractality strength, for differentiating between normal and malignant categories. In addition, it also exhibited the potential to conceptually correlate breast texture to BC molecular subtypes.

In mammograms, the detection of masses is challenging since normal and abnormal (or pre-cancerous and cancerous) tissues look similar, which escalates the emergence of false positives in computer-aided diagnostics. To overcome this limitation, mammograms facilitated five feature extraction methods were proposed, out of which two are Hilbert space-filling curve-based image representation and fractal texture analysis. [58] The authors argued that the extraction of features directly from the complete ROI overcomes the need for image segmentation and also takes into account lesions surrounding the tissues, which can be useful in BC diagnosis. Subsequently, another study utilized Hilbert curves to investigate a set of 111 mass contours for differentiating between 65 benign and 46 malignant masses. An accuracy of 99% was achieved in terms of the area under the receiver operating characteristic (ROC) curve. [59]

Nonetheless, while analyzing 2D images captures the spatial heterogeneity, it can be computationally expensive. Consequently, to reduce computational complexity and to emphasize boundary-related morphological features, an approach has been proposed by transforming 2D images into their 1D sequences or contour signatures for fractal analysis, and subsequently, the $FD$ of breast masses was computed using Higuchi's algorithm. The authors suggested the method to be easy and quick in implementation, and can serve as an auxiliary measure in pathological diagnosis. [60] In another work, although using Higuchi's algorithm, ultrasound-radio frequency time-series analysis was performed to classify malignant breast lesions. A machine learning framework combined with time-series features was used to generate malignancy maps for depicting the likelihood of malignancy within a region of 1 $mm^2$ containing suspicious lesions. The resulting ROC curve exhibited an accuracy of 86% (81%) at a 95% confidence interval from Support Vector Machines (Random Forest Classification) in combination with time-series features, and consequently, can reduce the number of unnecessary biopsies in mammography screening [61] for the early detection of BC.

Dynamic contrast-enhanced MRI is a robust technique for the diagnosis of BC in high-risk women, although similar contrast between benign masses and malignant lesions limits the sensitivity of the technique. In regard to the aforementioned, Soares *et al* [62] proposed a 3D MFA with $LC$ as the multifractal measure. The result suggested the effectiveness of the method in differentiating between benign and malignant samples as judged by the support vector machine classification method, with an accuracy of 96%. Another study investigated the applicability of 3D MFA for classifying benign and malignant breast tumors and for assessing chemotherapeutic response using dynamic contrast-enhanced MRI. The study enrolled twenty-four female patients between 18–60 years with a mean age of 45.0 ± 3.4 years, diagnosed with BC. The participants were divided into two cohorts, i.e., Group 1 (10 nos.), in which MRI breast images were analyzed to estimate multi-fractal scaling exponents with $LC$ as a supplementary measure; and Group 2 (14 nos.), in which MRI images obtained before and after chemotherapy were analyzed to evaluate treatment effectiveness. Image pre-processing was performed, including conversion of images to binary format to reduce data loss, morphological dilation to connect fragmented domains, and selection of a cubic ROI (32–64 pixels) according to the BI-RADS classification. Distinct $LC$ and $FD$ ranges were reported for the three studied groups. Malignant neoplasms exhibited the highest LC values (0.53–0.81), benign tumors showed intermediate values (0.19–0.42), and control tissues showed the lowest (0.05–0.09). Correspondingly, the $FD$ increased from controls (1.6–2.1) to benign (2.01–2.35) and malignant (2.47–2.81) lesions. ROC analysis indicated a diagnostic sensitivity (specificity) of 70.8% (65.9%) for the $FD$ with an area under the curve (AUC) = of 0.71 and 72.4% (89.3%) for LC with AUC = 0.84. The multi-fractal parameters also reflected treatment response. Across two chemotherapy sessions, MRI-measured tumor diameters decreased markedly. The mass exponent exhibited increased nonlinearity, while the singularity spectra $h(q)$ broadened after chemotherapy, signifying greater heterogeneity reduction and enhanced self-similarity of the tissue structure. The Hurst exponent, which was initially below 0.4, increased to 0.7–0.9 after treatment, suggesting the emergence of long-term spatial correlations consistent with therapy-induced tissue normalization. Finally, in the multi-fractal spectrum and from a biophysical standpoint, the left branch ($q > 0$) corresponded to regions of reduced blood supply and low signal intensity, whereas the right branch ($q < 0$) represented regions with increased perfusion and high intensity. Hence, the spectrum width served as an indicator of tumor vascular heterogeneity and was proposed as a quantitative descriptor of chemotherapeutic effectiveness. [63]

The reported studies demonstrate that fractal descriptors retain diagnostic relevance across imaging modalities, enabling quantification of morphological complexity that parallels histological observations.

## 3.3 Computational Morphodynamics and Stochastic Models

FG also serves as a theoretical foundation for modeling tumor growth and morphological evolution. Computational morphodynamics integrates stochastic rules, scaling hypotheses, and spatial self-affinity to simulate the irregular growth fronts characteristic of invasive tumors.

### 3.3.1 Computational Morphodynamics Modeling

It has been argued that intense vascularization is an essential condition for neoplastic development. [64] In this regard, the growth of solid neoplasms was proposed to accompany neovascularization, where the growth of new capillaries is more vigorous and continuous than the growth of capillary shoots in fresh wounds and inflammations. [65] A major feature in neovascularization is the growth factor, namely, endothelial cells, in tumour blood, which responds to angiogenic factors by upregulating the proliferation, migration, and differentiation rates while lowering the apoptosis rate. However, these effects are reported to be insufficient in explaining the vascular architecture in tumours. Also, in the absence of growth factor, as a consequence of the oxygen gradient in normal tissues, the tumour vasculature grows vigorously through a heterogeneous extracellular matrix by the process of invasion percolation.

Baish and Jain [66] used a computer model of tumour vasculature based on the process of percolation to investigate the transport of drugs and oxygen to tumours. They observed a lack of overall network optimization in tumour tissues when compared to normal tissues. Specifically, they noted that regressing of tumour, e.g., androgen-dependent Shionogi, modeled blood vessels after hormone removal display a notable fractal scaling (possibly, space-filling behaviour) similar to normal tissues, which indicates that a reduction in the number of vessels and consequently, vessel density may improve the transport of blood-borne substances to the tumour, which can promote tumour growth. Similar results were reported for flow correlated percolation during remodeling of vessels in growing solid tumours. [67] Accordingly, an emphasis should be given to the need for tumour investigation from an architectural and physiological perspective to aid in molecular diagnostic tools. Ribeiro *et al* [68] inspected the form and growth process of avascular tumours, considering the competition between cells. The proposed model reproduced the conventionally observed early-stage exponential growth followed by power-law growth resulting from the tumour's fractal structure, facilitated by the emergent optimal value of $FD$ as a consequence of interaction between cells. In addition, a similar relation was observed to hold between the intrinsic replication-competition rate between cells and the allocation of energy in growing animals. In other words, a universal behaviour was proposed to exist in the growth of avascular tumours and animals, which can be modelled using the Bertalanffy-Richards model. In a different work, d'Onofrio extended the mean-field theory proposed by Mombach *et al* [69], which aimed to mechanistically link macroscopic tumour properties to the microscopic cell properties, via focusing on the role of cell-cell interactions only. He concluded that the interaction of a cell with the microenvironment can be encoded in the form of noise-induced fluctuations by considering the parameters that take into account the proliferation rate of a cell, resulting from its baseline replication rate and a constant related to the effectiveness of inhibitory actions. Interestingly, these two parameters were integrated in a single equation by assuming a fractal spatial structure of cell populations. [70] This observation supports that quantification of morphological complexity in cell clusters, representing the spatial organization of cells, can provide useful diagnostic information.

### 3.3.2 Stochastic Morphodynamics Modeling

The stochastic interface models or dynamic scaling theory simulate tumor-front propagation based on local random growth rules. Tumour invasion involves nonlinear, stochastic front dynamics causing morphological instabilities and heterogeneous spread, beyond cell-autonomous models. In the framework of dynamic scaling theory, cancer cell colonies, expanding as clusters, show roughening consistent with Kardar–Parisi–Zhang (KPZ) scaling, from Eq. 23, while non-cancerous cells can show comparable roughening, indicating stochastic collective motion independent of malignancy. [71, 72] Also, tumour–host interfaces under radial symmetry display scaling consistent with KPZ or Edwards–Wilkinson (EW) classes. Models with surface diffusion and noise predict interface width, correlation length, and fractal geometry [73] with growth dynamics often between KPZ and molecular beam epitaxy (MBE), governed by Eq. 24 and characterized by super-rough interfaces, edge-constrained proliferation, and radial linear growth. [74] These mesoscopic- or tissue-scale laws classify fronts, reveal mechanisms, robustness, and morphological evolution, and can connect the microscopic (or cellular) and macroscopic (tumor) scale behavior and dynamics.

$$\frac{\partial h(x,t)}{\partial t} = \nu \nabla^2 h + \frac{\lambda}{2} (\nabla h)^2 + \eta(x,t) \tag{23}$$

$$\frac{\partial h(x,t)}{\partial t} = -K \nabla^4 h + F + \eta(x,t) \tag{24}$$

Here, $h(x,t)$ represents the local displacement of the tumour interface position at the lateral coordinate $x$ and time $t$ and interpreted as the local radial distance of the invasive front from the tumour centre; $\nu$ is the coefficient of surface tension; $\lambda$ captures the lateral growth asymmetry; $K$ and $F$ represent the coefficient of surface diffusion and growth rate; and $\eta(x,t)$ is a white or colored and zero-mean Gaussian additive noise term with variance: $\langle \eta(x,t)\eta(x',t') \rangle = 2D\delta(x-x')\delta(t-t')$. Notably, $\eta(x,t)$ encapsulates the aggregated effects of microscopic biological variability and local environmental heterogeneity. For example, it represents deviations from the average proliferative rate at the tumour boundary, arising from stochastic processes such as fluctuations in cell division timing, uneven distribution of biochemical cues, localized hypoxia, and random migratory behaviours.

The interface width or roughness $w(L,t)$ of the tumour interface of arc-length $L$ evolves according to a scaling function $w(L,t) \sim L^{\alpha} f\left(\frac{t}{L^z}\right)$ and computed using Eq. 23:

$$w(L,t) = \langle [h(x,t) - \langle h(x,t) \rangle]^2 \rangle^{1/2} \tag{23}$$

Subsequently, it is used to extract the Family-Vicsek scaling exponents, given from Eq. (24a-c):

$$\alpha: w(L, t \gg t_c) \sim L^{\alpha} \tag{24a}$$

$$\beta: w(L, t \ll t_c) \sim t^{\beta} \tag{24b}$$

$$z = \frac{\alpha}{\beta} \tag{24c}$$

Here, $t_c$ being the cross-over time and $\alpha$ denotes the roughness exponent, $\beta$ the growth exponent, and $z$ the dynamic exponent. These relationships capture the self-affine nature of tumor interfaces, quantitatively linking microscopic cellular activity to emergent macroscopic morphology. The computational findings confirmed that tumor growth interfaces exhibit kinetic roughening consistent with self-similar scaling laws.

The approach is supported by experimental data, as reported by Brú *et al* [74]. The authors applied scaling analysis to contours of 15 cell lines (grown *in-vitro*) and 16 tumour types (developed *in-vivo*), including breast adenocarcinoma and nodal metastases, respectively, and concluded that cell colony and tumor interfaces exhibit fractal characteristics with $FD = 1.31$ for the BC cases. In addition, they observed that the computed scaling exponents were observed to be compatible with the MBE universality class. This physical observation was supported experimentally by three orthogonal observations viz., (i) spatial mapping of proliferation (bromodeoxyuridine (BrdU) in cell colonies, Ki-67 in tumors) shows that proliferative activity concentrates in a narrow peripheral band (e.g., the outer 20% of area can contain most active cells, consistent with the spatial organization of tumour spheroid), (ii) time-lapse tracking demonstrates short-range movement of daughter cells along the interface toward regions of higher local curvature, and (iii) radius growth curves show an early transient followed by an approximately linear increase of mean radius with time in *in-vitro* colonies. Subsequently, they proposed that surface diffusion of newly born cells along the tumor border is the coarse-grained mechanism giving rise to observed roughening, and that tumor growth should be regarded primarily as a competition for space rather than for nutrients. Nevertheless, they explicitly acknowledge the limitations that temporal exponents ($\beta$ *and* $z$) could not be obtained for clinical tumors, and circular or expanding geometry may shift theoretical exponent values, so the *in-vivo* universality assignment remains plausible but not directly measured.

Dynamic scaling theory provides a bridge between static fractal morphometrics and spatiotemporal evolution, describing how spatial complexity develops over time through self-affine interface roughening. In biophysical terms, the scaling exponents correspond to the roughness of the tumor front, the rate of morphological evolution, and the correlation between spatial and temporal scales. These findings highlight the relevance of this approach in capturing the interplay between cellular proliferation, mechanical interactions, and morphological complexity, offering a theoretical dynamical framework complementary to the static image analyses presented in the preceding sections.

# 4. Fractional Calculus for Dynamical Analysis

## 4.1 Fundamentals of Fractional Operators and Growth Laws

FC provides a mathematical framework for describing complex biological processes that exhibit memory, non-locality, and heterogeneity in space and/or time. In biological tissues, such effects arise from structural complexity, viscoelastic interactions, and feedback extending across multiple scales. [21] Fractional operators generalize classical integer-order differentiation and integration to non-integer orders ($\alpha$) and introduce memory (temporal non-locality) or long-range spatial interactions via convolutional kernels. The rationale for using fractional operators in tumor and tissue modeling is that integer-order derivatives fail to represent these long-range temporal correlations and spatial couplings that are intrinsic to cell proliferation, nutrient diffusion, and stress-relaxation phenomena. Specifically, $\alpha$ quantifies the strength of memory or correlation in the system. When $\alpha \to 1$, the model converges to classical integer-order or Markovian dynamics, while smaller values reflect stronger non-local effects, associated with sub-diffusion or prolonged cellular response times. Thus, FC provides a continuum for modeling anomalous transport, delayed signaling, and possibly, self-organizing morphodynamics that cannot be represented within traditional calculus frameworks. In the following, we have discussed some mathematical preliminaries of FC, biological motivation and interpretation, applications specific to BC, and the conceptual correlation between FG and FC that can aid in BC research.

The foundational equation in FC is the Riemann–Liouville (RL) definition, which states that a fractional order integral of order $\Re(\alpha) > 0$ follows from Cauchy's formula for repeated integrals. [75] Subsequently, the Riemann–Liouville fractional integral is defined for dynamic systems from Eq. (25):

$$I^{\alpha} f(t) = \frac{1}{\Gamma(\alpha)} \int_0^t (t-\tau)^{\alpha-1} f(\tau) d\tau \tag{25}$$

Here, $t > 0, \alpha \in R^+, \Gamma(\alpha) = \int_0^{\infty} t^{\alpha-1} e^{-t} dt$ is the Euler-Gamma function, and $f(t)$ is a causal function of $t$.

The fractional-order Riemann–Liouville derivative is defined from Eq. (26):

$$D_{RL}^{\alpha} f(t) = \frac{d^m}{dt^m} \left[ \frac{1}{\Gamma(\mathrm{m}-\alpha)} \int_0^t \frac{f(\tau)}{(t-\tau)^{\alpha-m+1}} d\tau \right] \tag{26}$$

Here, $m \in \mathbb{N}$ and $m - 1 < \alpha < m$. Physically, these operators imply that the current rate of change depends on the entire prior history of $f(t)$ weighted by a singular power-law kernel $(t-\tau)^{\alpha-1}$, representing long-range memory.

An alternative definition of the fractional-order Riemann–Liouville derivative was introduced by Caputo (C), as defined by Eq. (27). [75] This definition has the advantage of incorporating both initial and boundary conditions and also overcomes the limitation of the Riemann–Liouville integral, which yields a non-zero value for the derivative of a constant for the initial-value problems. For example, the Riemann–Liouville fractional derivative of order $1/2$ with respect to the time, of a constant is $constant/\sqrt{\pi t}$.

$$D_{C}^{\alpha} f(t) = \frac{1}{\Gamma(\mathrm{m}-\alpha)} \int_0^t \frac{f^{m}(\tau)}{(t-\tau)^{\alpha-m+1}} d\tau \tag{27}$$

Nonetheless, the mentioned definitions utilized a singular kernel, which poses the risk that the integrand or the derivative can become infinite or undefined at some point in its domain. To overcome this limitation, Caputo and Fabrizio introduced a derivative with an exponential kernel, defined by Eq. (28a): [76]

$$D_{CF}^{\alpha}f(t) = \frac{M(\alpha)}{1-\alpha}\int_a^t f'(\tau)exp\left[-\gamma\frac{t-\tau}{1-\alpha}\right]d\tau \tag{28a}$$

Here, $M(\alpha)$ is so chosen that the operator recovers classical limits at $\alpha = 0,1$; where, $\alpha \in (0,1]$ is the fractional order. However, Eq. (28a) is only defined for $f(t) \in H^1(a,b), b > a$. In case, $f(t) \notin H^1(a,b)$, then the Caputo-Fabrizio (CF) differential operator is defined from Eq. (28b):

$$D_{CF}^{\alpha}f(t) = \frac{\alpha M(\alpha)}{1-\alpha}\int_a^t (f(t) - f(\tau))exp\left[-\gamma\frac{t-\tau}{1-\alpha}\right]d\tau \tag{28b}$$

The fractional integral in the Caputo-Fabrizio sense is defined from Eq. (29):

$$I_t^{\alpha}f(t) = \frac{2(1-\alpha)}{M(\alpha)(2-\alpha)}f(t) + \frac{2\alpha}{M(\alpha)(2-\alpha)}\int_0^t f(\tau)d\tau \tag{29}$$

Here, $t \geq 0$ and $\alpha \in (0,1)$. Also, the condition (Eq. 30) follows from Eq. (29).

$$\frac{2(1-\alpha)}{M(\alpha)(2-\alpha)} + \frac{2\alpha}{M(\alpha)(2-\alpha)} = 1 \tag{30}$$

$$\Rightarrow M(\alpha) = \frac{2}{2-\alpha}$$

This gives a modified form of the CF differential operator, defined by Eq. (31):

$$D_{CF}^{\alpha}f(t) = \frac{1}{1-\alpha}\int_a^t f'(\tau)exp\left[-\gamma\frac{t-\tau}{1-\alpha}\right]d\tau \tag{31}$$

The operator can be used to model complex systems with fading memory, i.e., the contribution of the past states decays exponentially rather than following a power-law, as was the case in Riemann–Liouville and Caputo's definitions. Notably, it can be used to model processes with finite memory, such as transient biochemical signaling.

The Atangana–Baleanu (AB) family further generalizes the kernel to the Mittag-Leffler function, which bridges exponential and power-law memory. It is to be noted that as the fractional order approaches one, the Mittag-Leffler function converges to the exponential solution. The AB differential operator in the Riemann–Liouville and Caputo sense is defined by Eq. (32) and Eq. (33), respectively. [76]

$$D_{ABRL}^{\alpha}f(t) = \frac{A(\alpha)}{1-\alpha}\frac{d}{dt}\int_a^t f(\tau)E_{\alpha}\left[-\frac{\alpha}{1-\alpha}(t-\tau)^{\alpha}\right]d\tau \tag{32}$$

$$D_{ABC}^{\alpha}f(t) = \frac{A(\alpha)}{1-\alpha}\int_a^t f'(\tau)E_{\alpha}\left[-\frac{\alpha}{1-\alpha}(t-\tau)^{\alpha}\right]d\tau \tag{33}$$

The fractional order integral in the Atangana–Baleanu sense is defined from Eq. (34):

$$I_t^{\alpha}f(t) = \frac{1-\alpha}{A(\alpha)}f(t) + \frac{\alpha}{A(\alpha)\Gamma(\alpha)}\int_0^t f(\tau)\,(t-\tau)^{\alpha-1}d\tau \tag{34}$$

Here, $E_{\alpha}[]$ is the Mittag-Leffler function and $A(\alpha)$ is a normalization constant with $A(0) = A(1) = 1$. A detailed discussion on FC in the context of its mathematics, including the existence and uniqueness of the solutions for the fractional equations (e.g., via Banach fixed-point theorems or Picard iterations), solutions (e.g., Laplace transform), stability and asymptotic (e.g., Ulam-Hyers stability), the available numerical approaches for solving the equations (e.g., Adams–Bashforth–Moulton predictor–corrector method, etc.), and dimensional consistency, along with its broad applications in bioengineering, is beyond the scope of this review. Nevertheless, interested readers are directed to references for accessible introductions to its concepts and applications. [19, 20, 76-80]

A prominent research area in mathematical oncology is modeling the growth of tumors. In this context, conventional phenomenological ordinary differential equations-based laws (exponential, Gompertz, Bertalanffy) fit tumour volume retrospectively without a mechanistic grounding in tumour biology [81] and ignore spatial structural heterogeneity, tumour-microenvironment interactions, stochasticity and memory, failing to understand the underlying system [82-83] and to capture complexities like invasive front instability, morphological irregularity, and post-treatment relapse. Cancer malignancies arise from stochastic, uncontrolled proliferation that evades apoptosis, producing solid tumours (carcinomas, sarcomas, lymphomas), which account for nearly 90% of cases, including breast, colon, prostate, bladder, and lung cancers. [84] While genetic mechanisms are well studied, tumour growth and invasion remain poorly understood. [81] Growth reflects not only cancer cell accumulation via the so-called hallmarks of cancer, but also emergent collective dynamics, spatial constraints, nutrient diffusion, immune responses, mechanical feedback, heterogeneity, and memory. Tumour progression is thus a stochastic, nonlinear, multi-scale process rather than a simple increase in cell number. Capturing these dynamics is essential for prognosis, therapy, drug scheduling, and early detection.

As highlighted above, fractional order operator-derived models include memory and subsequently, fractional order counterparts of the conventional growth laws, e.g., fractional order exponential model, logistic model, Gompertz model, and Bertalanffy-Putter model were defined, as given from Eq. (35a-d), respectively. [75]

$$\frac{d^\alpha v}{dt} = a \cdot v(t) \tag{35a}$$

$$\frac{d^\alpha v}{dt} = a \cdot v(t) \cdot \left(1 - \left(\frac{v(t)}{K}\right)^b\right) \tag{35b}$$

$$\frac{d^\alpha v}{dt} = a \cdot v(t) \cdot ln\left(\frac{b}{v(t)+c}\right) \tag{35c}$$

$$\begin{cases} \frac{d^\alpha v}{dt} = p \cdot v^a - q \cdot v^b, for\ a \neq b \\ \frac{d^\alpha v}{dt} = p \cdot v^a - \ln(v) \cdot q \cdot v^a, for\ a = b \end{cases} \tag{35d}$$

Here, $v$ is the tumour volume, $a$ is the kinetic parameter or inherent growth rate, $b$ is the correction factor to the growth rate, $K$ is the average population of a species (e.g., volume of cancerous cells), $c$ is the minimum volume carrying capacity, $p$ is the intrinsic growth, and $q$ is the growth rate of the anti-angiogenic process.

## 4.2 Application of Fractional Calculus in Breast Cancer Research

The avascular growth phase is a critical and significant part of the growth of solid tumors, including the breast. As discussed earlier, researchers have studied the form and growth process of avascular tumours, considering the competition between cells. [68] In this regard, a breast cancer competition model was modeled by Solís-Pérez *et al* [85] considering the population dynamics among healthy, cancer stem, and tumour cells, respectively, along with the effect of excess estrogen and the body's natural immune response on the cell populations using Liouville-Caputo and Caputo-Fabrizio-Caputo fractional derivatives. A numerical scheme was employed using the Laplace transform (Atangana-Toufik) to obtain the special numerical solutions. The results suggested the positive significance of fractional derivatives for revealing the complexity of dynamics in the proposed model. Valentim Jr. *et al* [83] investigated the significance of FC for tumor growth prediction in 6- to 8-week-old nude mice in which GI-101A human BC cells were introduced and established as xenografts using fractional growth laws. Subsequently, they attributed the superior performance of FC in the Caputo sense over integer order calculus (via Goodness-of-fit indicators) to the incorporation of memory in the model, since tumours accumulate mutations and other variations during their evolution. In another study using the same data, the authors investigated the multi-stage tumor characteristics using a variable-order $\alpha(t)$ fractional equation-based model in the Caputo sense, whose solution was obtained in a variable-order Mittag–Leffler form. Notably, $\alpha(t)$ was interpreted as an index of memory where $\alpha \approx 1$ represents the memoryless exponential growth, while $\alpha < 1$ implied stronger non-local or long-range memory effects. Optimization using the global search minimizing the sum of squared residuals, shows progressive improvement with higher-order polynomials and a periodic $\alpha(t)$ providing the best fit ($R^2$ up to 0.9969), suggesting oscillatory memory dynamics or alternating faster (slower) growth, possibly representing dormancy (activation) cycles or shifts between cancer hallmarks. Numerical predictor–corrector verification using a modified Adams scheme confirms the approximate analytical solution within 0.4% error for the tested profiles. [86] This study demonstrates that variable-order fractional models can substantially improve descriptive accuracy over classical and fixed-order fractional exponentials and argues for future testing of variable $\alpha(t)$ in other tumor types and alternative growth laws. In addition, while the aforementioned study focused on capturing multistage tumor evolution through a time-dependent fractional order that modulates intrinsic memory, a different study [87] extended the fractional approach toward modeling the systemic consequences of treatment, which integrates fractional operators into a chemotherapy–cardiotoxicity framework for BC dynamics. Here, the authors proposed a fractional-order compartmental model to capture the interplay between BC progression, chemotherapy, and its cardiotoxic effects. The model partitions the patient population into five compartments viz., early-, intermediate-, and advanced-stage cancer, recovered, and cardiotoxic groups, and employs the CF fractional derivative to account for finite memory with exponential decay. The existence and uniqueness of the obtained solutions were proven via fixed-point theory and Lipschitz contraction, and numerical integration was performed using the Adams–Bashforth scheme, which shows that reducing the fractional order slows tumor dynamics, reflecting stronger memory and delayed treatment response. Parametric sweeps demonstrate that higher recovery rates increase disease-free population but also cardiotoxicity, while increased cardiotoxic transition rates rapidly amplify cardiac complications. Comparisons with the integer-order system confirm that fractional dynamics yield smoother, biologically realistic temporal profiles. Conceptually, the model captures chemotherapy-induced systemic feedback within a finite-memory framework. Nonetheless, the study is phenomenological and lacks explicit spatial or cellular detail. Another study [88] modeled cycle-specific chemotherapy by replacing classical time derivatives with the Caputo fractional derivative to account explicitly for memory effects in proliferating $(P)$ and quiescent $(Q)$ tumor cells. The model employed the compartmentalization approach and utilized the two compartments, i.e., proliferating, treatment-sensitive, and quiescent, treatment-resistant, and numerical solutions were obtained using a piecewise on–off type function. The integer-order system was solved analytically for each on/off interval using matrix exponentials and characteristic multipliers, and subsequently, served as the baseline for comparison. The fractional generalization, written with Caputo derivative of order $\alpha \in (0,1]$, is transformed into equivalent integral equations and solved numerically using finite-difference approximations for both the Caputo derivative and the Riemann–Liouville integral. Specifically, the authors studied how the maximum characteristic multiplier and the resulting conditions for net decay vs. growth depend on active drug time, period, and dose strength, and how the fractional order modifies trajectories of proliferating and quiescent cell masses, and the proliferative function, which represents treatment effects, and defined as $r(t) = \frac{P(t)}{P(t)+Q(t)}$. Numerically, it was observed that a smaller $\alpha$ or stronger memory smooths transitions and can produce decay behavior during drug-off periods; not visible from the integer order modeling. Consequently, the authors argue that appropriately choosing order and active drug time could inform cycle scheduling. Additionally, the study reports the existence of analytic solutions and derives stable numerical schemes for the modeled fractional system, along with presenting parameter studies illustrating the memory effects.

Interestingly, FC has also found application in image analysis. Specifically, in image analysis, the point at which brightness changes sharply or has a discontinuity indicates edges. The detection of edges decreases the amount of data to be processed and filters redundant information while preserving the structural information. Lavín-Delgado *et al* proposed an FC-based method in the Caputo sense for improving the edge-detection and consequently, contrast and texture of mammograms for easy detection of microcalcifications and images from other body parts. [76] They showed the significant performance of the fractional operator over existing methods for the detection of both boundaries and edges, respectively, and argued that the approach not only maintains the low-frequency features of contours in the smooth regions, but also enhances high-frequency components (e.g., edges, textures).

The discussed studies, although limited in numbers, highlight the possible augmentation in BC research regarding solid tumor dynamics, therapy resistance, and early detection and/or diagnosis.

### 4.3 Integrating Fractal Geometry and Fractional Calculus in an Integrative Morphological Framework for Breast Cancer: A Conceptual Perspective

FG and FC are complementary approaches for investigating the multi-scale complexity and heterogeneity in the complex BC system. Here, FG quantifies spatial, mesoscopic organization and dynamics using fractal parameters and dynamic scaling theory, whereas FC captures non-local temporal memory and anomalous transport via fractional operators. West [78], in his perspective, discussed the possible integration of FG and FC for the description of complex human physiology. Specifically, it was reported that applying a fractional derivative of order $\alpha$ to a fractal function of dimension $FD$ produces a new fractal function with dimension $FD + \alpha$, highlighting a biophysically interpretable bridge between a spatial fractal measure ($FD$) and a temporal fractional operator of order ($\alpha$). In another work [89], he demonstrated that stochastic or time-varying fractional orders can generate multifractal dynamics, i.e., randomness or temporal dependence in fractional order can produce multifractal time series. These observations conceptually imply that $\alpha$ and $FD$ are related and that fractional order can be interpreted as a contribution to the fractal structure or behavior of complex systems.

In the context of fractal morphometrics in image analysis and/or dynamic scaling, it can be hypothesized that if the case of a tissue or tumor front has a defined $FD$, then a fractional operator with order $\alpha$ chosen so that the model outputs reflect a specific/target dimension in accordance with the reported perspective. Nonetheless, it should be treated as an interpretative mapping since, to the best of our knowledge, there is no literature reporting an estimation procedure of fractional order from the $FD$ of an image. Also, observation of multi-fractal signatures in data may indicate the need for models with a time-varying fractional order. This implies that in cases where 2D-MFDFA reveals broad multifractal spectra, FC should be treated as the modeling approach to account for the complexity and/or heterogeneity in the investigated system.

The readers should note that this review discusses the conceptual correlation or integration between FG and FC and not between fractal calculus and FC, which is beyond the scope of this review. However, the discussion can be considered as a conceptual rationale for the fractal-fractional modeling approach of complex systems, which can take into account both local and non-local behavior in space and time, as introduced by Atangana [90] and its application to tumor growth dynamics in general and concerning BC, reported in the literature. [91-93] In addition, the integration is theoretically supported since FG descriptors can and should inform FC model choice and parametrization because FC operators mathematically relate to fractal function properties, and fractional order can generate multifractal behavior.

## 4.4 Strengths, Limitations, and Best Practices

FC provides a mechanistic modeling approach for temporal evolution, non-local transport (e.g., anomalous diffusion), treatment response with memory, or to study dynamics where past states matter. The fractional order and chosen kernels determine the system memory and anomalous scaling in governing equations. The parameters (order, kernels) are estimated by fitting dynamical models to temporal data or cumulative observations, which also offers the limitation of under-determination without rich datasets. Explicit description regarding specific utilized operator(s), initial or boundary conditions, existence and uniqueness of solutions, their stability, numerical techniques, growth or dynamic regimes, computational complexity, tolerance, etc., should be shared for reproducibility and biophysically meaningful interpretation.

Nonetheless, it should be noted that these are model parameters that govern dynamics and are estimated in the context of a chosen model, not directly measured from images. Though there is a direct conceptual correlation between FG and FC, there is currently no mathematically consistent procedure for directly incorporating FD from image or dynamic scaling analysis in fractal-fractional modeling of complex biological systems like BC.

## 5. Future Prospects and Challenges

The convergence of FG and FC has significant potential in enhancing the morphometric and dynamical understanding of complex BC systems. However, the review of existing literature indicates that substantial methodological, computational, and translational challenges remain unresolved before these frameworks can mature into reproducible, clinically meaningful tools. Future research should prioritize expanding the dimensional and methodological scope of current FG–FC analyses and establishing rigorous standards for data acquisition, parameter estimation, and validation. As noted here, one immediate direction lies in extending mono-fractal 2D studies toward 3D and multi-fractal paradigms capable of capturing the hierarchical tissue architecture and heterogeneity of tumor microenvironments. This shift will not only improve the geometric representation of normal, benign, and malignant tissues but also allow the integration of fractal morphological descriptors with fractional-order models describing non-local diffusion and memory-dependent growth processes. Furthermore, small-sample FD approaches demonstrated in limited cytological and haematological studies may be refined and standardized for low-cost, minimally invasive diagnostic applications in BC, offering translational potential in resource-constrained settings.

The principal methodological challenges concern data dimensionality, preprocessing sensitivity, and parameter identifiability. The majority of existing FG studies rely on mono-fractal analyses of mammograms or histological sections, which fail to reflect volumetric complexity and the multi-fractal nature of tumor evolution. Similarly, fractal parameter estimates are highly sensitive to pre-processing factors such as thresholding, ROI selection, and image resolution. These dependencies can introduce non-stationarity and reduce reproducibility across independent datasets. Also, standardization of preprocessing protocols, including fixed ROI dimensions, uniform thresholding criteria, and the explicit reporting of imaging metadata, has been repeatedly emphasized as an indispensable step toward method reproducibility. In the FC approach, the over-parameterization of fractional-order models presents another critical obstacle. Consequently, it should be acknowledged that inclusion of excessive parameters, especially when involving multiple fractional kernels and derivative definitions, can lead to biologically inconsistent results. Transparent justification for the selected fractional operator and the rationale for fractional order values is therefore essential. Computational tractability is another concern, since numerical evaluation of fractional operators and multi-fractal measures demands substantial resources, with finite-size and edge effects contributing additional biases if uncorrected. Beyond algorithmic aspects, a persisting gap lies in connecting mathematical descriptors to biological interpretation: while $FD$, multifractal strength, or $\alpha$ variations are measurable, their direct mechanistic correspondence to biophysical processes such as epithelial–mesenchymal transition, angiogenesis, or metastatic dissemination is seldom established, underscoring the need for integrative theoretical and experimental frameworks.

Future investigations must equally focus on the creation and dissemination of benchmark datasets to evaluate the generalizability of FG–FC approaches across imaging modalities, equipment types, and sample size or patient populations. Currently, most reported studies are retrospective, single-center investigations, often lacking independent validation cohorts. Multi-center and prospective validation studies, supported by harmonized metadata standards, are urgently required. The reviewed modeling literature further emphasizes that reproducibility cannot be ensured without the release of open-source analysis pipelines and numerical codes. Publicly accessible repositories containing simulation scripts, parameter files, and representative datasets would enable verification and reproducibility.

Translating FG–FC methodologies into clinical practice introduces distinct practical and ethical considerations. While current FG-based classifiers achieve promising diagnostic accuracies, their integration into radiological or pathological workflows requires higher sensitivity and specificity benchmarks to attain clinical relevance. Combining fractal descriptors with structural, molecular, or textural features represents a promising avenue for improving diagnostic performance while maintaining interpretability. Furthermore, to ensure clinical adoption, FG and FC-derived metrics must demonstrate operational compatibility with existing imaging systems and pathology infrastructures. According to us, the greatest translational opportunity lies in leveraging these mathematically driven methods to generate quantitative, explainable biomarkers from small and inexpensive samples, a pathway particularly advantageous in limited-resource healthcare contexts. Nonetheless, the ethical deployment of fractional models mandates transparent validation and a stepwise verification process prior to prognostic or therapeutic use.

From the aforementioned discussions, a coherent research roadmap can be delineated. Specifically, priority should be given to the standardization and open dissemination of FG–FC analysis pipelines. The benchmark datasets encompassing mammography, MRI, and histopathology with detailed acquisition metadata should be assembled to enable multi-center cross-validation and sensitivity testing. Also, hybrid biomarkers that integrate FG-based morphometric measures with clinical and molecular data should be designed and validated across independent cohorts to evaluate diagnostic and prognostic value. Finally, the integration of FG characterization with FC modeling frameworks, e.g., coupling measured vascular $FD$ with fractional diffusion models of nutrient transport, can be pursued to provide mechanistically interpretable insights into tumor invasion and therapeutic response.

Finally, despite the growing conceptual and methodological clarity of FG and FC in cancer systems, their translation to standardized and clinically robust applications remains hindered by several persistent challenges, as discussed in this work. The future of this research area depends on resolving these limitations through standardized reporting, open data and code sharing, rigorous cross-validation, and enhanced biological interpretability. Once these methodological and infrastructural gaps are addressed, FG–FC frameworks are poised to evolve into a unified quantitative paradigm capable of capturing both the spatial heterogeneity and temporal memory intrinsic to the BC system.

## 6. Conclusion

This review aims to consolidate evidence and subsequently demonstrate that integrating FG FC provides a quantitative approach for describing the multiscale structural and dynamical complexity of BC. FG-based descriptors such as $FD$, $LC$, $SC$, and multifractal spectra effectively capture morphological irregularities across cytological, histological, and radiological domains, while FC-based formulations incorporate memory and non-local dynamics relevant to tumor growth and anomalous transport. Collectively, these approaches extend beyond classical models by linking spatial heterogeneity with the temporal evolution of cancer systems.

The existing literature highlights that FG parameters can discriminate between normal, benign, and malignant tissues and reveal multiscale textural signatures in mammograms, histopathology, and MRI. In addition, FC operators, such as the Riemann–Liouville and Caputo derivatives, offer a rigorous framework for modeling diffusion and growth processes with fractional order as a tunable parameter controlling system memory. However, a direct mathematical correspondence between image- or dynamic scaling theory-derived fractal descriptors and fractional-order parameters remains conceptual rather than established.

Nonetheless, the comparative assessment and mentioned challenges highlight critical limitations like dependence on image preprocessing, predominance of 2D mono-fractal analyses, lack of standardized reporting, over-parameterization in FC models, and insufficient cross-dataset validation. Addressing these issues requires harmonized preprocessing protocols, open-source implementations, and benchmark datasets across imaging modalities. Developing reproducible pipelines and sensitivity analyses for parameter selection will be essential for clinical translation.

To summarize, FG and FC provide complementary, mechanistically consistent tools for quantifying BC morphology and dynamics. Their combined application holds potential for reproducible, interpretable biomarkers and integrative growth models once methodological and validation gaps are resolved. Realizing this integration will advance FG–FC frameworks from promising analytical constructs to reliable quantitative instruments for BC characterization and systems-level modeling.

**Acknowledgment: A. D.** acknowledges the Department of Biotechnology (DBT)-India for the Research Associate fellowship vide Award Letter No. DBT-RA/2023/January/NE/3594.

**R. B.** acknowledges support from the Indo-French Centre for the Promotion of Advanced Research (69T08-2).

**M. K. J.** acknowledges support from the Param Hansa Philanthropies.

**Declaration of Interest:** The authors declare that they have no known competing financial interests or personal relationships that could have appeared to influence the work reported in this paper.